\documentclass[10pt, letterpaper, oneside]{article}
\usepackage[utf8]{inputenc}
\usepackage[margin=1in]{geometry}
\usepackage{textcomp}
\usepackage{graphicx}
\usepackage{placeins}
\usepackage{amsfonts}
\usepackage{amsmath}
\usepackage{mathtools}
\usepackage{upgreek}
\usepackage{gensymb}
\usepackage{tabularray}
\usepackage[round]{natbib}
\usepackage[onehalfspacing]{setspace}
\usepackage[colorlinks=true, allcolors=black, citecolor=blue]{hyperref}
\usepackage{lineno}
\usepackage[format=plain, labelfont=it, textfont=it]{caption}
\usepackage{pdflscape}

\title{
	\textbf{Spherical convolutional neural networks can improve brain microstructure estimation from diffusion MRI data}
}

\author{
	Leevi Kerkelä$^1$*, Kiran Seunarine$^{1,2}$, Filip Szczepankiewicz$^3$, and Chris A. Clark$^1$
}

\date{
	\begin{flushleft}
	\scriptsize{
	    $^1$ UCL Great Ormond Street Institute of Child Health, University College London, London, United Kingdom \\
		$^2$ Great Ormond Street Hospital, London, United Kingdom \\
		$^3$ Medical Radiation Physics, Clinical Sciences Lund, Lund University, Lund, Sweden \\
		[2ex]
		* Corresponding author: leevi.kerkela.17@ucl.ac.uk; Developmental Imaging \& Biophysics Section, UCL Great Ormond Street Institute of Child Health, 30 Guilford Street, WC1N 1EH, London, United Kingdom
	}
    \end{flushleft}
}

\begin{document}

\maketitle

\section*{Abstract}

Diffusion magnetic resonance imaging is sensitive to the microstructural properties of brain tissue. However, estimating clinically and scientifically relevant microstructural properties from the measured signals remains a highly challenging inverse problem that machine learning may help solve. This study investigated if recently developed rotationally invariant spherical convolutional neural networks can improve microstructural parameter estimation. We trained a spherical convolutional neural network to predict the ground-truth parameter values from efficiently simulated noisy data and applied the trained network to imaging data acquired in a clinical setting to generate microstructural parameter maps. Our network performed better than the spherical mean technique and multi-layer perceptron, achieving higher prediction accuracy than the spherical mean technique with less rotational variance than the multi-layer perceptron. Although we focused on a constrained two-compartment model of neuronal tissue, the network and training pipeline are generalizable and can be used to estimate the parameters of any Gaussian compartment model. To highlight this, we also trained the network to predict the parameters of a three-compartment model that enables the estimation of apparent neural soma density using tensor-valued diffusion encoding.

\section*{Keywords}

Diffusion magnetic resonance imaging; Geometric deep learning; Microstructure; Spherical convolutional neural network

\thispagestyle{empty}

\newpage


\section{Introduction}

Neuroimaging enables non-invasively measuring functional and structural properties of the brain, and it is essential in modern neuroscience. Diffusion magnetic resonance imaging (dMRI), the most commonly used imaging modality for quantifying microstructural properties of the brain, measures displacements of water molecules at the microscopic level and is thus sensitive to tissue microstructure. dMRI has been used to localize microstructural alterations associated with, for example, learning \citep{sagi2012learning}, healthy development \citep{lebel2019review}, ageing \citep{sullivan2006diffusion}, neurodevelopmental disorders \citep{gibbard2018structural}, and neurodegenerative diseases \citep{zhang2009white}. However, accurately inferring clinically and scientifically relevant properties of tissue microstructure (e.g., cell morphology or distribution of cell types) from the measured signals remains a highly challenging inverse problem \citep{kiselev2017fundamentals}.

Most dMRI data analysis methods are based on signal models that express the measured signal as a function of parameters of interest and can be fit to data by numerically minimizing an objective function \citep{novikov2019quantifying}. An essential requirement for microstructural neuroimaging methods is low rotational variance (i.e., estimated parameters should not depend on how the subject's head is oriented in the scanner). Furthermore, it is often desirable for the parameter estimates to be independent of the orientation distribution of the microscopic structures (e.g., an estimate of axon density should not depend on whether the axons are aligned or crossing). These two requirements are often achieved by acquiring high-angular resolution diffusion imaging (HARDI) data and averaging over the diffusion encoding directions, which is referred to as "powder-averaging", a term borrowed from the field of solid-state nuclear magnetic resonance (NMR). The number of acquisition directions required for a nearly rotationally invariant powder-averaged signal depends on the properties of tissue microstructure and diffusion encoding \citep{szczepankiewicz2019tensor}. Fitting models to powder-averaged signals is often referred to as the "spherical mean technique" (SMT), a term introduced by \cite{kaden2016quantitative}. While powder-averaging enables the estimation of various microstructural parameters \citep{jespersen2013orientationally, lasivc2014microanisotropy, kaden2016quantitative, kaden2016multi, szczepankiewicz2016link, henriques2020correlation, palombo2020sandi, gyori2021potential}, a significant amount of information is lost during averaging. Therefore, it may be beneficial to estimate the parameters directly from full data without powder-averaging.

In recent years, microstructural parameter estimation using machine learning (ML) has received significant attention as a potential solution to issues with conventional fitting, such as slow convergence, poor noise robustness, and terminating at local minima \citep{golkov2016q, barbieri2020deep, palombo2020sandi, karimi2021deep, gyori2021potential, de2021neural, elaldi2021equivariant, sedlar2021diffusion, sedlar2021spherical, gyori2022training, kerkela2022improved}. ML models can be trained to predict microstructural parameter values from data using supervised or self-supervised learning. In the context of dMRI, a particularly promising development has been the invention of spherical convolutional neural networks (sCNNs) \citep{cohen2018spherical, esteves2018learning}. sCNNs are $\text{SO}(3)$-equivariant (i.e., rotating the input changes the output according to the same rotation) artificial neural networks that perform spherical convolutions with learnable filters. They theoretically enable rotationally invariant classification and regression, making them potentially well-suited for predicting microstructural parameters from dMRI data.

This study aimed to investigate if sCNNs can improve microstructural parameter estimation. We focused on estimating the parameters of a constrained two-compartment model by \cite{kaden2016multi} regularly used in neuroscience to study human white matter \textit{in vivo} \citep{collins2019white, voldsbekk2021sleep, toescu2021tractographic, rahmanzadeh2022new, battocchio2022bundle}. An sCNN implemented according to \cite{esteves2018learning} was trained to predict the neurite orientation distribution function (ODF) and scalar parameters (neurite diffusivity and density) from dMRI data. Training and testing were done using simulated data. The sCNN was compared to conventional fitting and a multi-layer perceptron (MLP) in terms of accuracy and orientational variance. The trained model was then applied to MRI data acquired in a clinical setting to generate microstructural maps. Furthermore, to highlight the fact that the sCNN and training pipeline are applicable to any Gaussian compartment model, the network was trained to estimate the parameters of a constrained three-compartment model by \cite{gyori2021potential} that enables the estimation of apparent neural soma density using tensor-valued diffusion encoding \citep{topgaard2017multidimensional}.

\section{Materials and methods}

\subsection{Spherical harmonics}

Any square-integrable function on the sphere $f: S^2 \rightarrow \mathbb{C}$ can be expanded in the spherical harmonic basis:
\begin{equation}\label{eq:isft}
f(\mathbf{x}) = \sum_{l=0}^b \sum_{m=-l}^l \hat{f}_l^m Y_l^m(\mathbf{x}) ,
\end{equation}
where $\mathbf{x}$ is a point on the unit sphere, $b$ is the bandwidth of $f$, $l$ is the degree, $m$ is the order, $\hat{f}_l^m$ is an expansion coefficient, and $Y_l^m$ is a spherical harmonic defined as
\begin{equation}
Y_l^m(\theta, \phi) = \sqrt{\frac{2l+1}{4\pi}\frac{(l-m)!}{(l+m)!}} P_l^m(\cos \theta) e^{im\phi} ,
\end{equation}
where $\theta \in \left[ 0, \pi \right]$ is the polar coordinate, $\phi \in \left[0, 2 \pi \right)$ is the azimuthal coordinate, and $P_l^m$ is the associated Legendre function.

The expansion coefficients are given by the spherical Fourier transform (SFT):
\begin{equation}\label{eq:sft}
\hat{f}_l^m = \int_{S^2} \text{d}\mathbf{x} \ f(\mathbf{x}) \bar{Y}_l^m(\mathbf{x}) .
\end{equation}
SFT of a band-limited function can be computed exactly as a finite sum using a sampling theorem \citep{driscoll1994computing}. Equation \ref{eq:isft} is the inverse spherical Fourier transform (ISFT).

Since reconstructed dMRI signals are real-valued and antipodally symmetric, we use the following basis:
\begin{equation}\label{eq:basis}
S_l^m = \left\{\begin{matrix*}[l]
0 & \text{if } l \text{ is odd} \\
\sqrt{2} \ \Im \left( Y_l^{-m} \right) & \text{if} \ m < 0 \\ 
Y_l^0 & \text{if} \ m = 0 \\
\sqrt{2} \ \Re \left( Y_l^{m} \right) & \text{if} \ m > 0 
\end{matrix*}\right . .
\end{equation}

Considering that diffusion encoding directions do not usually follow a sampling theorem like the one by \cite{driscoll1994computing} that enables SFT to be exactly computed as a finite sum, we use least squares to compute the expansion coefficients: Indexing $j = \frac{1}{2}l(l + 1) + m$ assigns a unique index $j$ to every pair $l, m$. Given $f$ sampled at points $\mathbf{x}_1, \mathbf{x}_2, ..., \mathbf{x}_{n_\text{points}}$ stored in a column vector $\mathbf{X}$, the values of the spherical harmonics sampled at the same points are organized in a $n_\text{points} \times n_\text{coefficients}$ matrix $\mathbf{B}$ where $B_{ij} = S_l^m(\mathbf{x}_i)$. $\left( \mathbf{B}^{\text{T}}\mathbf{B} \right)^{-1} \mathbf{B}^{\text{T}} \mathbf{X}$ gives a vector containing the expansion coefficients minimizing the Frobenius norm \citep{brechbuhler1995parametrization}.

\subsection{Spherical convolution}

Convolution of a spherical signal $f$ by a spherical filter $h$ is defined as
\begin{equation}\label{eq:sph_conv}
(f * h)(\mathbf{x}) = \int_{\text{SO}(3)} \text{d}\mathbf{R} \ f(\mathbf{R} \mathbf{\hat{e}_3}) h(\mathbf{R}^{-1} \mathbf{x}) , 
\end{equation}
where $\mathbf{\hat{e}_3}$ is a unit vector aligned with the $z$-axis. If $f$ and $h$ are band-limited, the above equation can be evaluated efficiently as a pointwise product in the frequency domain \citep{driscoll1994computing}. The spherical harmonic coefficients of the convoluted signal $y$ are
\begin{equation}\label{eq:sph_conv_sh}
\hat{y}_l^m = 2 \pi \sqrt{\frac{4 \pi}{2l + 1}} \hat{f}_l^m \hat{h}_l^0 .
\end{equation}

Spherical convolution is equivariant to rotations (i.e., $\mathbf{R}(f * h) = (\mathbf{R}f) * h$ for all $\mathbf{R} \in \text{SO}(3)$) and the filter is marginalized around the $z$-axis (i.e, for every $h$, there exists a filter $h_z$ that is symmetric with respect to the $z$-axis so that $f * h = f * h_z$).

\subsection{Compartment models}

Compartment models represent the dMRI signal as a sum of signals coming from different microstructural environments (e.g., intra- and extra-axonal water). For details, see, for example, the review by \cite{jelescu2017design}. Here, we focus on models with non-exchanging Gaussian compartments following an ODF. The signal measured along $\mathbf{\hat{n}}$ is expressed as a spherical convolution of the ODF by a microstructural kernel response function $K$: 
\begin{equation}\label{eq:model_signal}
S(\mathbf{\hat{n}}) = \int_{\text{SO}(3)} \text{d}\mathbf{R} \ \text{ODF}(\mathbf{R} \mathbf{\hat{e}_3}) K(\mathbf{R}^{-1} \mathbf{\hat{n}}) ,
\end{equation}
where $K$ is the microstructural kernel response function:
\begin{equation}\label{eq:kernel_signal}
K(\mathbf{\hat{n}}) = S_0 \left[ \sum_{i=1}^N f_i \exp(- \mathbf{b}:\mathbf{D}_i) \right] ,
\end{equation}
where $S_0$ is the signal without diffusion-weighting, $N$ is the number of compartments, $f_i$ is a signal fraction, $\mathbf{b}$ is the b-tensor corresponding to $\mathbf{\hat{n}}$ and a b-value equal to $\text{Tr}(\mathbf{b})$, $:$ denotes the generalized scalar product ($\mathbf{b}:\mathbf{D} = \sum_{i=1}^3 \sum_{j=1}^3 b_{ij} D_{ij}$) \citep{westin2016q}, and $\mathbf{D}_i$ is an axially symmetric diffusion tensor aligned with the $z$-axis representing Gaussian diffusion in the compartment. The training pipeline presented in this paper is applicable to any compartment model that can be expressed using equations \ref{eq:model_signal} and \ref{eq:kernel_signal}. Given a different data generation method, the sCNN can be trained to predict the parameters of non-Gaussian models as well.

\subsubsection{Two-compartment model}

The so-called "standard model" of diffusion in white matter consists of a one-dimensional compartment representing diffusion inside neurites and a coaxial axially symmetric extra-cellular compartment \citep{novikov2019quantifying}. We focus on a constrained version of the model by \cite{kaden2016multi} that enables model parameters to be estimated from powder-averaged data using the SMT. The model contains two parameters: intra-neurite diffusivity $d$ and intra-neurite signal fraction $f$. Axial and radial diffusivities of the extra-cellular compartment are $d$ and $(1 - f)d$, respectively. Inserting this into Equation \ref{eq:kernel_signal} gives

\begin{equation}\label{eq:two-compartment_kernel_signal}
 K(\mathbf{\hat{n}}) = S_0 \left[ f \exp\left(- \mathbf{b}:\begin{bmatrix} 0 & 0 & 0 \\ 0 & 0 & 0 \\ 0 & 0 & d \end{bmatrix}\right) + (1-f) \exp\left(- \mathbf{b}:\begin{bmatrix} (1-f)d & 0 & 0 \\ 0 & (1-f)d & 0 \\ 0 & 0 & d \end{bmatrix}\right) \right] .
\end{equation}

\subsubsection{Spherical mean technique}

\cite{kaden2016quantitative} observed that for a fixed b-value, the spherical mean of the dMRI signal over the gradient directions does not depend on the ODF. By exploiting this invariance, the constrained two-compartment model can be fit to powder-averaged data, denoted by $S_\text{PA}$ here, using the following signal equation \citep{kaden2016multi}:

\begin{equation}
S_\text{PA} = S_0 \left[ f \frac{\sqrt{\pi}\text{erf}\left( \sqrt{bd}\right)}{2\sqrt{bd}} + (1-f)e^{-b(1-f)d} \frac{\sqrt{\pi}\text{erf}\left( \sqrt{bfd}\right)}{2\sqrt{bfd}}\right] .
\end{equation}

\subsubsection{Three-compartment model}
 
\cite{palombo2020sandi} added a spherical compartment representing neural soma to the standard model to make it more suitable for gray matter. We use a constrained three-compartment model by \cite{gyori2021potential} that uses tensor-valued diffusion encoding to make apparent neural soma imaging more feasible without high-performance gradient hardware. The model contains four parameters: intra-neurite diffusivity $d_\text{i}$, intra-neurite signal fraction $f_\text{i}$, spherical compartment diffusivity $d_\text{sph}$, and spherical compartment signal fraction $f_\text{sph}$. Axial and radial diffusivities of the extra-cellular compartment are $d_\text{i} (1 - f_\text{i} - f_\text{sph})^{\frac{1}{2} f_\text{sph} / (f_\text{sph} + f_\text{i})}$ and $d_\text{i} (1 - f_\text{i} - f_\text{sph})^{(\frac{1}{2} f_\text{sph} + f_\text{i}) / (f_\text{sph} + f_\text{i})}$, respectively. We omit explicitly writing out the kernel signal equation to save space, but it is trivial to construct from Equation \ref{eq:kernel_signal}.

\subsection{Simulations}

Simulated training data was generated by evaluating Equation \ref{eq:model_signal} in the frequency domain according to Equation \ref{eq:sph_conv_sh}. The response function values were evaluated along 3072 directions uniformly distributed over the surface of the sphere according to the hierarchical equal area isolatitude pixelisation (HEALPix) \citep{gorski2005healpix, zonca2019healpy} and expanded in the spherical harmonics basis. Rician noise was added to the simulated signals:
\begin{equation}
S_{\text{noisy}} = \sqrt{ \left( S + X \right)^2 + Y^2} ,
\end{equation}
where $S$ is the simulated signal without noise and $X$ and $Y$ are sampled from a normal distribution with zero mean and standard deviation of $1/\text{SNR}$, where SNR is the signal-to-noise ratio. SNR was matched to the mean SNR in the imaging experiments.

\subsection{Network architecture}

\begin{figure}
  \centering
  \includegraphics[width=.74\linewidth]{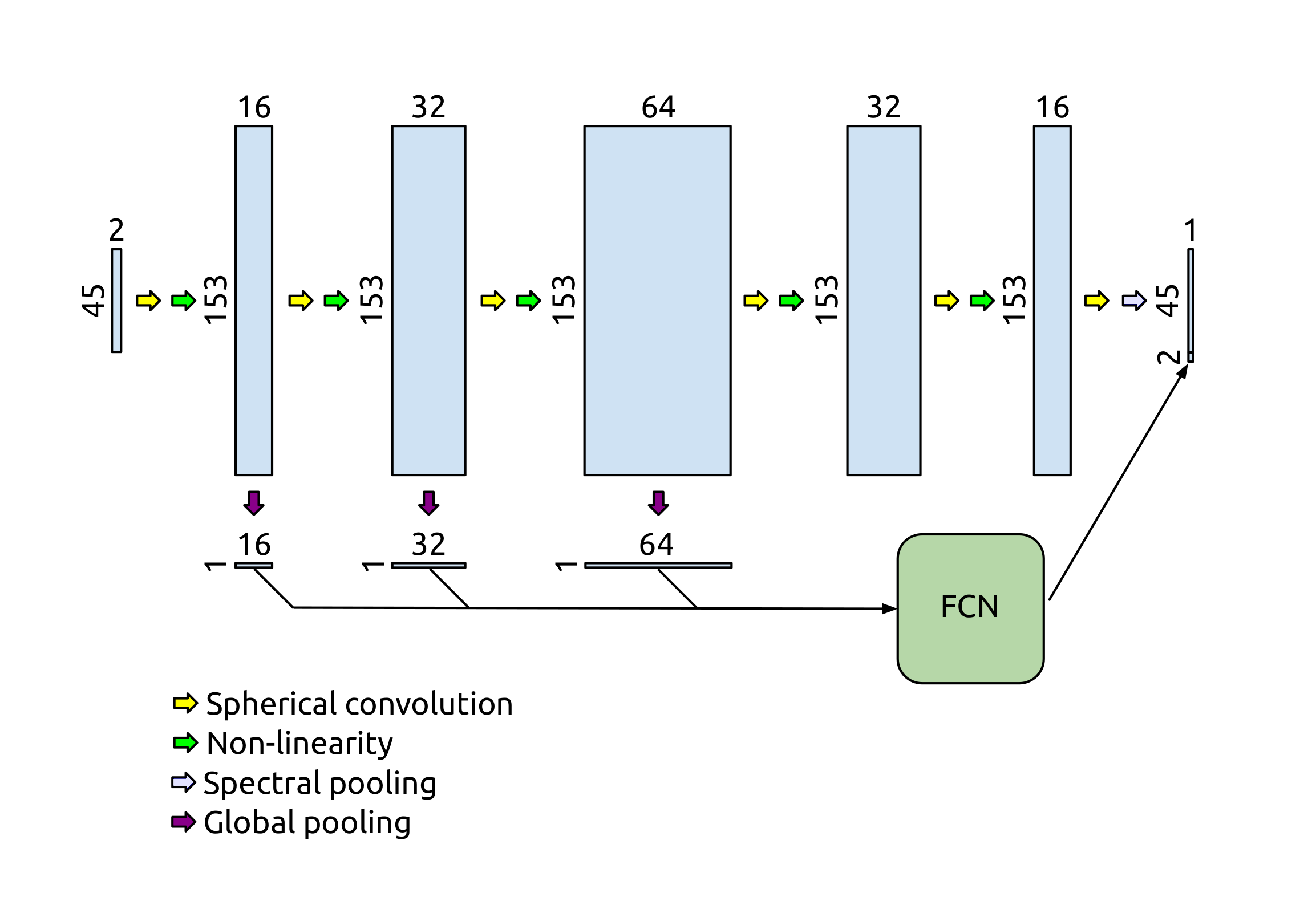}
  \caption{Network for two-compartment model parameter prediction. The input is normalized two-shell data expanded using spherical harmonics up to degree eight. The signals undergo spherical convolutions, non-linearities, and spectral pooling to produce the predicted orientation distribution function. After the initial three convolutions, global mean pooling is applied in the signal domain, and the resulting arrays are concatenated to create a nearly rotationally invariant feature vector passed on to the FCN that outputs the predicted scalar parameter.}
  \label{fig:architecture}
\end{figure}

Our sCNN, visualized in Figure \ref{fig:architecture}, consists of six spherical convolution layers implemented according to \cite{esteves2018learning} without enforcing localized filters. The network takes the expansion coefficients in the frequency domain as input and outputs the estimated ODF and scalar model parameters. The number of input channels is equal to the number of shells in data. Each spherical convolution layer is followed by a leaky (slope is 0.1 for negative values) rectified linear unit (ReLU) applied in the spatial domain. The conversion between frequency and spatial domains is done using the 3072 HEALPix directions. Spherical harmonics up to degree 16 are used in the network because the non-linearity can increase signal bandwidth. Spectral pooling discards coefficients of the highest degrees. After the initial three convolutions, global mean pooling is applied in the spatial domain, and the resulting arrays are concatenated and passed to the fully connected network (FCN) that outputs the predicted scalar parameter. The FCN consists of three hidden layers with 128 units each. The first two layers of the FCN are followed by batch normalization \citep{ioffe2015batch} and a ReLU. The sCNN for estimating the two-compartment model parameters has 78,258 trainable parameters.
 
\subsection{Training}

Training was done over $10^5$ batches of simulated data generated during training. Each batch contained signals from 500 microstructural configurations produced by random sampling ($d \sim \text{U}(0, 3$ \textmu m$^2$/ms$)$ and $f \sim \text{U}(0, 1)$). ODFs were sampled from five volunteer scans. Validation and test datasets were constructed similarly, except that they contained 10$^4$ and 10$^6$ microstructural configurations, respectively, and the ODFs were sampled from different volunteer scans. Training was performed twice: with and without randomly rotating the ODFs. The ODFs in the validation and test datasets were randomly rotated. ADAM \citep{kingma2014adam} was the optimizer with an initial learning rate of 10$^{-3}$, which was reduced by 90\% after 50\% and 75\% into the training. Mean squared error (MSE) was the loss function. ODF MSE was calculated in the spatial domain.

\subsection{Baseline methods}

The sCNN was compared to the SMT and an MLP that takes the normalized dMRI signals as inputs and outputs the spherical harmonic coefficients of the ODF and the model parameters. The SMT parameter estimation and the subsequent ODF estimation using the estimated microstructural kernel and constrained spherical deconvolution (CSD) was done using Dmipy \citep{fick2019dmipy}. The MLP consisted of three hidden layers with 512 nodes each. The hidden layers were followed by batch normalization and a ReLU. The MLP had 614,447 trainable parameters. It was trained like the sCNN, except ten times more batches were used to account for the higher number of parameters and ensure convergence.

\subsection{Imaging data}

The brains of eight healthy adult volunteers were scanned on a Siemens Magnetom Prisma 3T (Siemens Healthcare, Erlangen, Germany) at Great Ormond Street Hospital, London, United Kingdom. Data was denoised \citep{veraart2016denoising} using MRtrix3 \citep{tournier2019mrtrix3} and distortion- and motion-corrected using FSL \citep{andersson2016integrated, jenkinson2012fsl}. SNR was estimated in each voxel as the inverse of the standard deviation of the normalized signals without diffusion-weighting.

\subsubsection{High-angular resolution diffusion imaging}

Seven volunteers were scanned using a standard clinical two-shell HARDI protocol with two non-zero b-values of 1 and 2.2 ms/\textmu m$^2$  with 60 directions over half a sphere each. Other relevant scan parameters were the following: diffusion time ($\Delta$) = 28.7 ms; diffusion encoding time ($\updelta$) = 16.7 ms; echo time (TE) = 60 ms; repetition time (TR) = 3,050 ms; field of view (FOV) = 220 × 220 ms; voxel size = 2 × 2 × 2 mm$^3$; slice gap = 0.2 mm; 66 slices; phase partial Fourier = 6/8; multiband acceleration factor = 2. Fourteen images were acquired without diffusion-weighting, one of which had the phase encoding direction reversed to be used to correct for susceptibility-induced distortions. The total scan time was 7 minutes. Mean SNR in the brain was 50. Neurite ODFs were estimated using multi-tissue CSD \citep{jeurissen2014multi} with $l_\text{max} = 8$.

\subsubsection{Tensor-valued diffusion imaging}

One volunteer was scanned using a prototype spin echo sequence that enables tensor-valued diffusion encoding \citep{szczepankiewicz2019tensor}. Data was acquired using numerically optimized \citep{sjolund2015constrained} and Maxwell-compensated \citep{szczepankiewicz2019maxwell} gradient waveforms encoding linear and planar b-tensors. The acquisitions with linear b-tensors were performed with b-values of 0.5, 1, 2, 3.5, and 5 ms/\textmu m$^2$ with 12, 12, 20, 20, and 30 directions over half a sphere, respectively. The acquisitions with planar b-tensors were performed with b-values of 0.5, 1, and 2 ms/\textmu m$^2$ with 12, 12, and 20 directions over half a sphere, respectively. Other relevant scan parameters were the following: TE = 82 ms; TR = 4.2 s; FOV = 220 × 220 ms; voxel size = 2 × 2 × 2 mm$^3$; slice gap = 0.2 mm; 66 slices; phase partial Fourier = 6/8; multiband acceleration factor = 2. Fourteen images were acquired without diffusion-weighting, one of which had the phase encoding direction reversed. The total scan time was 12 minutes. Mean SNR in the brain was 29.

\section{Results}

\FloatBarrier

\subsection{Two-compartment model}

\subsubsection{Prediction accuracy}

\begin{figure}
  \centering
  \includegraphics[width=.99\linewidth]{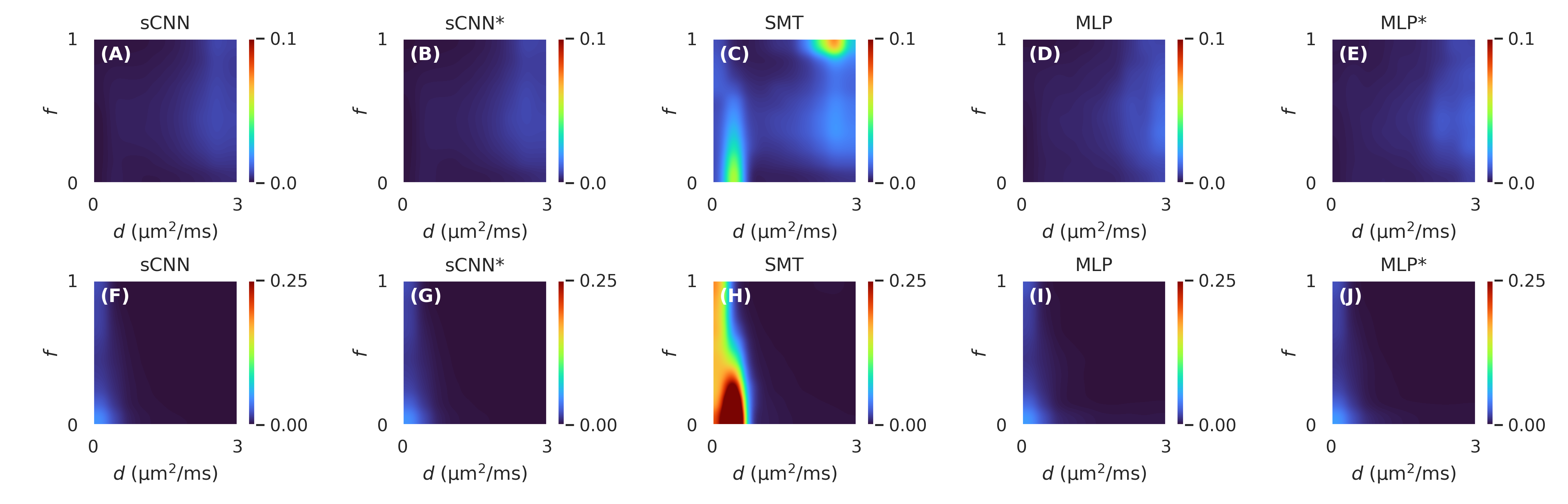}
  \caption{Mean squared error of the estimated two-compartment model parameters on the test dataset for different values of intra-neurite diffusivity ($d$) and intra-neurite signal fraction ($f$). The first row (A-E) shows the results for $d$ and the second row (F-J) shows the results for $f$. Deep learning-based methods outperformed the spherical mean technique in all parts of the parameter space. The asterisk (*) refers to models trained with randomly rotated training data.}
  \label{fig:2-compartment_model_prediction_accuracy}
\end{figure}

MSE on the test dataset is reported in Table \ref{tab:2-compartment_model_prediction_accuracy}. The sCNN and MLP outperformed the SMT in estimating the ODF and scalar parameters. The sCNN predicted $d$ and $f$ the best while the MLP was predicted the ODF marginally better than the sCNN. Both the sCNN and MLP benefited slightly from randomly rotating the training data. Figure \ref{fig:2-compartment_model_prediction_accuracy} shows how prediction accuracy depends on the values of $d$ and $f$. The sCNN and MLP outperformed the SMT in all parts of the parameter space. Although the largest errors with the SMT occurred for values of $d$ and $f$ not typically observed in the brain, ML-based approaches were more accurate for values observed in the brain (i.e., $d$ roughly between 1 and 2 \textmu m$^2$/ms).

\begin{table}
\centering
\begin{tblr}{ccc} 
\textbf{Method} & ODF & $d$ (\textmu m$^2$/ms) & $f$ \\
\hline
sCNN & $2.76 \cdot 10^{-3}$ & $3.08 \cdot 10^{-3}$ & $\mathbf{3.23 \cdot 10^{-3}}$ \\
sCNN* & $2.75 \cdot 10^{-3}$ & $\mathbf{3.07 \cdot 10^{-3}}$ & $\mathbf{3.23 \cdot 10^{-3}}$ \\
SMT & $6.47 \cdot 10^{-3}$ & $10.92 \cdot 10^{-3}$ & $37.50 \cdot 10^{-3}$ \\
MLP & $2.71 \cdot 10^{-3}$ & $4.00 \cdot 10^{-3}$ & $3.70 \cdot 10^{-3}$ \\
MLP* & $\mathbf{2.70 \cdot 10^{-3}}$ & $4.00 \cdot 10^{-3}$ & $3.63 \cdot 10^{-3}$
\end{tblr}
\caption{Mean squared error of the estimated two-compartment model parameters on the test dataset. Deep learning-based parameter estimation outperformed the spherical mean technique. The asterisk (*) refers to models trained with randomly rotated training data. The lowest values are highlighted in bold.}
\label{tab:2-compartment_model_prediction_accuracy}
\end{table}

\FloatBarrier

\subsubsection{Rotational variance}

The rotational variance of the different methods was assessed by generating signals from $10^3$ random microstructural configurations rotated over 729 rotations given by the SO(3) sampling theorem by \cite{kostelec2008ffts}. No noise was added to the signals to exclude the effects of noise. The average standard deviation of the estimated parameters from the rotated data are shown in Table \ref{tab:rotational_variance}. The sCNN and SMT were much less sensitive to rotations than the MLP. The SMT had the lowest rotational variance for $d$, and the sCNN had the lowest rotational variance for $f$. However, the SMT's non-zero rotational variance was driven by low values of $d$ or $f$ for which the fit is unstable. For values typically observed in white matter, the SMT's estimates' standard deviation was three orders of magnitude smaller than the average. Data augmentation by rotating the input signals improved prediction accuracy for both the sCNN and MLP. However, the sCNN was much less rotationally variant even without data augmentation than the MLP was with data augmentation.

\begin{table}[h!]
\centering
\begin{tblr}{ccc} 
\textbf{Method} & $d$ (\textmu m$^2$/ms) & $f$ \\
\hline
sCNN & $0.23 \cdot 10^{-3}$ & $0.13 \cdot 10^{-3}$ \\
sCNN* & $0.18 \cdot 10^{-3}$ & $\mathbf{0.09 \cdot 10^{-3}}$ \\
SMT & $\mathbf{0.14} \cdot 10^{-3}$ & $0.25 \cdot 10^{-3}$ \\
MLP & $20.30 \cdot 10^{-3}$ & $14.40 \cdot 10^{-3}$ \\
MLP* & $17.23 \cdot 10^{-3}$ & $12.78 \cdot 10^{-3}$
\end{tblr}
\caption{Average standard deviation of the estimated two-compartment model parameters over rotations of the input signals. The asterisk (*) refers to models trained with randomly rotated training data. The lowest values are highlighted in bold.}
\label{tab:rotational_variance}
\end{table}

\FloatBarrier

\subsubsection{Application on real imaging data}

\begin{figure}
  \centering
  \includegraphics[width=.99\linewidth]{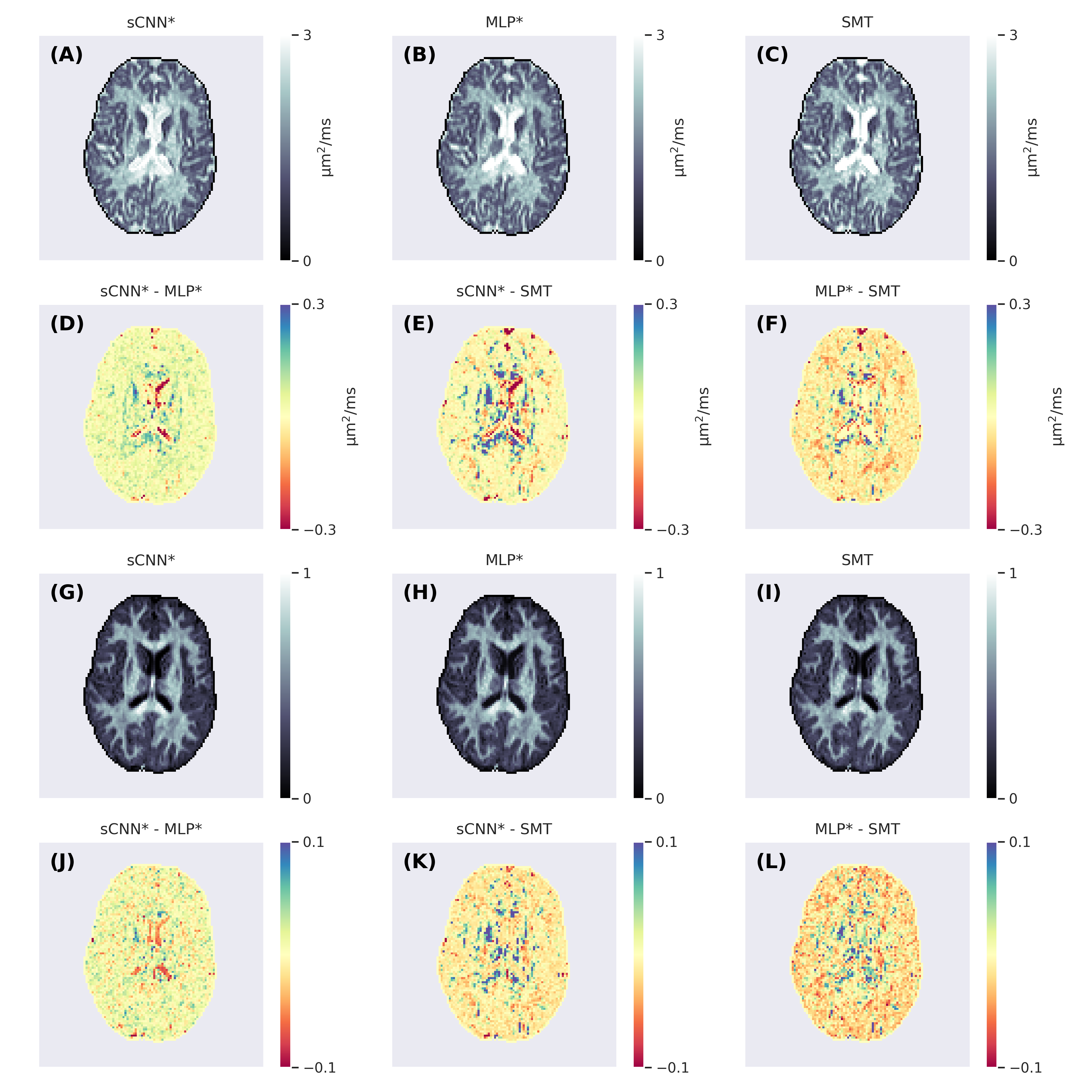}
  \caption{Axial slices of the intra-neurite diffusivity (A-C) and intra-neurite signal fraction (G-I) maps generated using the spherical convolutional neural network, multi-layer perceptron, and spherical mean technique. The second row (D-F) shows the differences between the intra-neurite diffusivity maps and the fourth row (J-L) shows the differences between the intra-neurite signal fraction maps.}
  \label{fig:2-compartment_model_maps}
\end{figure}

\begin{figure}
  \centering
  \includegraphics[width=.99\linewidth]{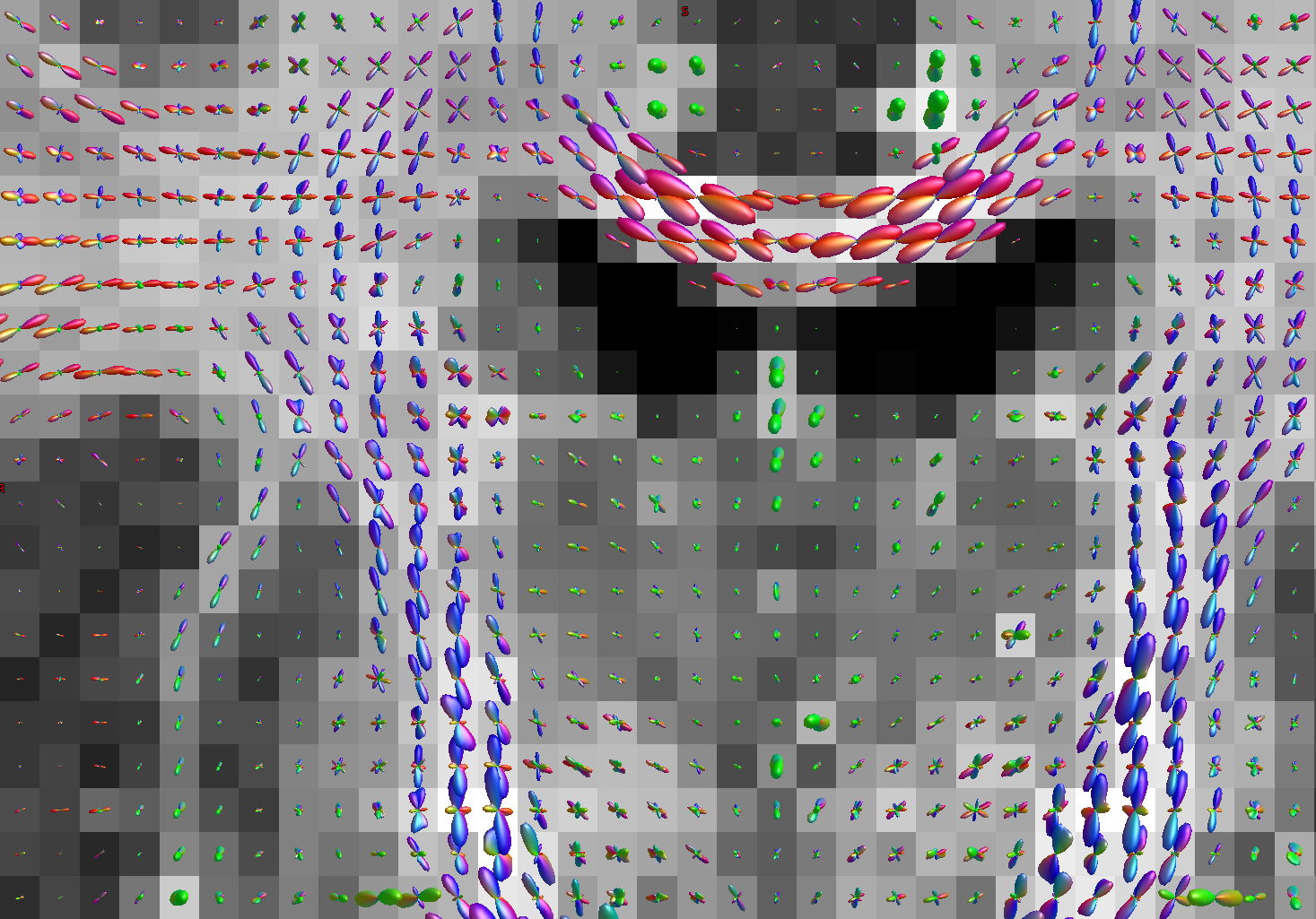}
  \caption{Neurite orientation distribution functions overlaid on a map of intra-neurite signal fraction generated by the spherical convolutional neural network trained with randomly rotating the training data. The colour represents the principal direction, and the size is scaled according to neurite density. This coronal slice shows the intersection of the corticospinal tract and the corpus callosum.}
  \label{fig:odfs}
\end{figure}

Figure \ref{fig:2-compartment_model_maps} shows parameter maps generated using the three methods. The maps produced by the ML-based methods appear less noisy. Overall, the sCNN estimated $d$ to be greater than the MLP (mean difference = $2.4 \cdot 10^{-2}$ \textmu m$^2$/ms; std of difference = $8.1 \cdot 10^{-2}$ \textmu m$^2$/ms) and SMT (mean difference = $0.9 \cdot 10^{-2}$ \textmu m$^2$/ms; std of difference = $12.7 \cdot 10^{-2}$ \textmu m$^2$/ms). However, in the CSF the sCNN tended to estimate $d$ to be less than the MLP or SMT. Overall, the sCNN estimated $f$ to be greater than the MLP (mean difference = $0.5 \cdot 10^{-2}$; std of difference = $3.6 \cdot 10^{-2}$) and SMT (mean difference = $0.1 \cdot 10^{-2}$; std of difference = $4.5 \cdot 10^{-2}$) while exhibiting a similar yet lesser tissue-dependent pattern as $d$. Figure \ref{fig:odfs} shows example ODFs generated by the trained sCNN.

\FloatBarrier

\subsection{Three-compartment model}

\begin{figure}
  \centering
  \includegraphics[width=.99\linewidth]{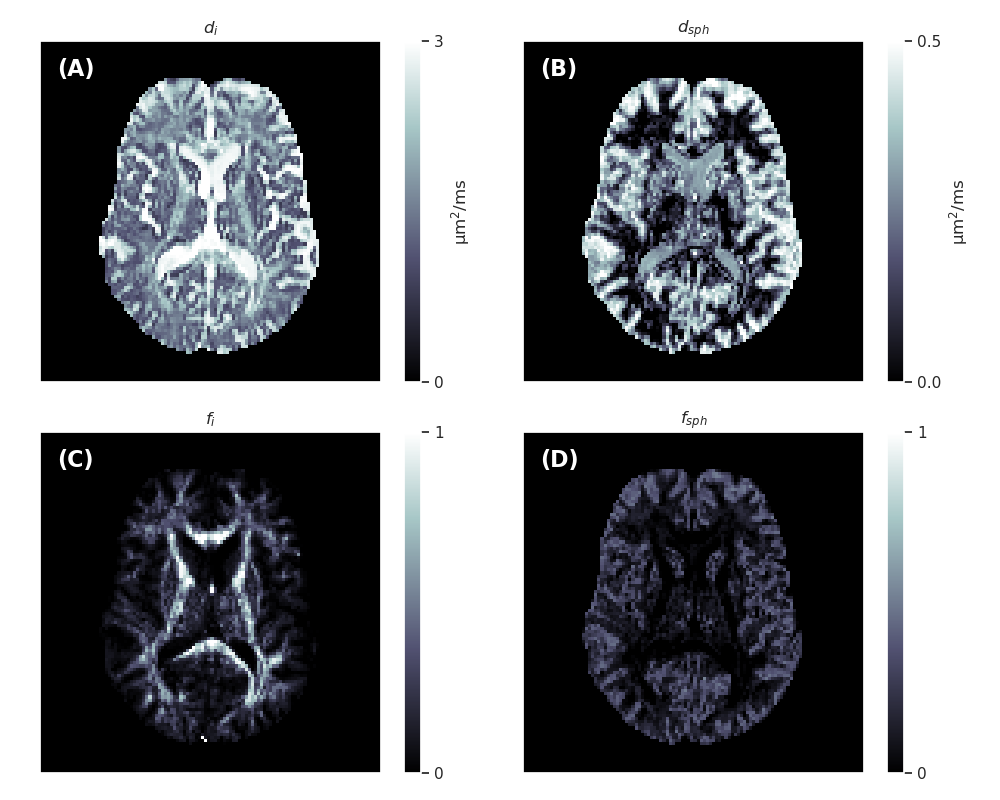}
  \caption{Axial slices of the intra-neurite diffusivity ($d_\text{i}$), spherical compartment diffusivity ($d_\text{sph}$), intra-neurite signal fraction ($f_\text{i}$), and spherical compartment signal fraction ($f_\text{sph}$) maps generated by the spherical convolutional neural network trained with randomly rotating the training data.}
  \label{fig:3-compartment_model_maps}
\end{figure}

To highlight the fact that the network and training pipeline are applicable to any Gaussian compartment models, the sCNN was trained to predict the three-compartment model parameters the same way as with the two-compartment model. Informed by the two-compartment model results, the network was trained with randomly rotated training data. $d_\text{i} \sim \text{U}(0, 3$ \textmu m$^2$/ms$)$, $f_\text{i} \sim \text{U}(0, 1)$, $d_\text{sph} \sim U(0, \text{max}(d_\text{i}, 0.5$ \textmu m$^2$/ms)), and $f_\text{sph} \sim \text{U}(0, f_\text{i})$. The upper limit of $d_\text{sph}$ was chosen to correspond to a sphere with a diameter of 25 \textmu m using the Monte Carlo simulator Disimpy \citep{kerkela2020disimpy}. Figure \ref{fig:3-compartment_model_maps} shows maps that the sCNN generated from preprocessed dMRI data.

\FloatBarrier

\section{Discussion}

The primary purpose of this study was to investigate whether sCNNs can improve microstructural parameter estimation from noisy dMRI data, focusing on a constrained two-compartment model widely used in neuroscience research to study human white matter \textit{in vivo}. The sCNN demonstrated superior accuracy with similar rotational variance compared to the SMT, and exhibited similar accuracy but considerably lower rotational variance than the MLP that had significantly more trainable parameters. Our results show that sCNNs can offer substantial benefits over simpler artificial neural network architectures for ML-based microstructural parameter estimation from dMRI data.

We focused on comparing neural network architectures with a fixed training strategy, using the SMT as a baseline. Previous research by \cite{gyori2022training} has highlighted the significant impact of training data distribution on neural network predictions, which affects the performance of our sCNN when applied to real imaging data. We are aware of this limitation, and in future work, we aim to optimize the training data distribution. Another relevant key takeaway from the work by \cite{gyori2022training} is that at low SNR, ML-based parameter estimation can suffer from high bias, which manifests as maps that appear exceedingly smooth. Moreover, it is important to note the general limitation of microstructural models that deviations from model assumptions can lead to inaccuracies \citep{lampinen2017neurite, henriques2019microscopic, kerkela2021comparative}.

When it comes to training the sCNN, while it is crucial to sample the space of possible ODFs as exhaustively as possible during training, the MLP training requirements are even more demanding since its rotational variance can only be reduced through learning. Changes in b-values or the angular resolution of shells will necessitate retraining our network. Technically, the same network could be used as long as the b-values remain consistent, but the spherical harmonics expansion would vary with different angular resolutions (i.e., the number of b-vectors).

To the best of our knowledge, sCNNs have been used to analyze dMRI data only a few times prior to this. \cite{sedlar2021spherical} trained an sCNN to predict 'neurite orientation dispersion and density imaging' (NODDI) \citep{zhang2012noddi} parameters from subsampled data, and \cite{goodwin2022can} showed that sCNNs can improve the robustness of diffusion tensor estimation from data with just a few directions. sCNNs have also been used to estimate ODFs \citep{sedlar2021diffusion, elaldi2021equivariant}. However, this study differs from the aforementioned studies in two important ways. First, our network and simulations were developed to estimate both the ODF and scalar parameters of any Gaussian compartment model. Second, we carefully compared the sCNN to the SMT, a commonly used and nearly rotationally invariant conventional fitting method, thus warranting a comparison with sCNN. Although we implemented spherical convolution layers as described by \cite{esteves2018learning}, other architectures also exist and warrant investigation in the context of microstructural parameter estimation. For example, the sCNNs by \cite{cohen2018spherical} use cross-correlation and can learn non-zonal (i.e., not symmetric with respect to the z-axis) filters, \cite{kondor2018clebsch} developed efficient quadratic nonlinearities in the spherical harmonics domain, and the graph-based sCNN by \cite{perraudin2019deepsphere} is suitable for spherical data with very high angular resolution. Besides optimizing network architecture, future studies should also focus on optimizing hyperparameters and especially on carefully assessing the effects of and optimizing the training data distribution.

\newpage

\section*{Data and code availability}

The code is publicly available at \url{https://github.com/kerkelae/scnn-investigation/}. Data may be shared for a reasonable request.

\section*{Author contributions}

LK developed the idea, acquired data, wrote the software, performed data analysis, and wrote the manuscript. KS contributed to data acquisition. FS wrote the code required for tensor-valued data acquisition. CC contributed to funding acquisition. All authors reviewed and approved of the manuscript prior to publication.

\section*{Conflicts of interest}

FS is an inventor on a patent related to the study.

\newpage

\bibliography{main}
\bibliographystyle{apalike}

\end{document}